\documentclass[]{jkas} 

\def\year{2024} 
\def\volume{??} 
\def\issue{?} 
\def\beginpage{1} 
\def\received{May 20, 2024} 
\def\accepted{June 20, 2024} 
\def\published{???? ??, 2024} 
\date{Received \received; Accepted \accepted; Published \published}

\title{%
Impact of Internal Dust Correction on the Stellar Populations of Galaxies Estimated Using the Full Spectrum Fitting
}

\author[1]{Joon Hyeop Lee}{0000-0003-3451-0925}
\author[1]{Hyunjin Jeong}{0000-0002-0145-9556}
\author[1]{Jiwon Chung}{0000-0003-0469-345X}
\author[2,3]{Mina Pak}{0000-0002-5896-0034}
\author[4]{Sree Oh}{0000-0002-4731-9604}

\affil[1]{Korea Astronomy and Space Science Institute, Daejeon 34055, Republic of Korea}
\affil[2]{School of Mathematical and Physical sciences, Macquarie University, Sydney, NSW 2109, Australia}
\affil[3]{ARC Centre of Excellence for All Sky Astrophysics in 3 Dimensions (ASTRO 3D), Australia}
\affil[4]{Department of Astronomy and Yonsei University Observatory, Yonsei University, Seoul 03722, Republic of Korea}

\def\corrauthor{%
J. H. Lee, \email{jhl@kasi.re.kr}
}

\def\runningauthor{%
Lee et al.
}

\def\runningtitle{%
Dust Correction in the Full Spectrum Fitting
}

\def\keywords{%
galaxies: abundances --- galaxies: stellar content
}

\def\abstracttext{%
Full spectrum fitting is a powerful tool for estimating the stellar populations of galaxies, but the fitting results are often significantly influenced by internal dust attenuation. For understanding how the choice of the internal dust correction method affects the detailed stellar populations estimated from the full spectrum fitting, we analyze the Sydney-Australian Astronomical Observatory Multi-object Integral field spectrograph (SAMI) galaxy survey data using the Penalized PiXel-Fitting (PPXF) package. Three choices are compared: (Choice-1) using the PPXF \emph{reddening} option, (Choice-2) using the multiplicative Legendre polynomial, and (Choice-3) using none of them (no dust correction). In any case, the total mean stellar populations show reasonable mass-age and mass-metallicity relations (MTR and MZR), although the correlations appear to be strongest for Choice-1 (MTR) and Choice-2 (MZR). When we compare the age-divided mean stellar populations, the MZR of young ($< 10^{9.5}$ yr $\approx 3.2$ Gyr) stellar components in Choice-2 is consistent with the gas-phase MZR, whereas those in the other two choices hardly are. On the other hand, the MTR of old ($\geq 10^{9.5}$ yr) stellar components in Choice-1 seems to be more reasonable than that in Choice-2, because the old stellar components in low-mass galaxies tend to be relatively younger than those in massive galaxies. Based on the results, we provide empirical guidelines for choosing the optimal options for dust correction.
}

\begin{document}
\jkashead 


\section{Introduction\label{intro}}

A galaxy's spectrum in the optical wavelength reveals a wealth of information about the galaxy, including stellar populations, gas content, kinematics of stars and gas, and internal dust attenuation. Among several approaches to extract refined information from the observed spectrum, which is usually a complex mixture of multiple ingredients of information, the full spectrum fitting \citep[FSF; e.g.,][]{gon14,che16,bar18,bar20,joh21,vau22,pak23} is a powerful method to estimate the stellar populations in the galaxy. Compared to other methods \citep[e.g., absorption line strength fitting;][]{pro04,sch07,tho11,mcd15,pak19,pak21,pak23}, the FSF has an advantage in its capability to estimate the history of star formation and chemical evolution as a function of lookback time, as well as the total mean stellar population. However, the FSF also has a weakness in that it is challenging to precisely measure the reliability or quantitative uncertainty of the estimated stellar populations.

\citet{lee23} is one of the rare studies that measured the statistical uncertainty of the stellar populations estimated using the FSF. By comparing a large sample of artificially composed galaxy spectra (input) with their FSF results (output), \citet{lee23} tabularized how the uncertainty of the estimated stellar age and metallicity depends on the signal-to-noise ratio (S/N) and the luminosity fraction of the stellar components in a given age interval, which are critical factors that determine the uncertainty of the stellar populations. However, there is another important element that may significantly affect the FSF results: internal dust attenuation. Dust tends to make the optical spectrum of a galaxy redder than its pure stellar components really are \citep[so-called blanketing effect;][]{mil28}, which may lead to biases in the estimated stellar populations, making them appear older or more metal-rich than they actually are.

To correct for the effects of internal dust, we may adopt a specific function of dust attenuation \citep[e.g., Calzetti law;][]{cal00} or simply adopt a high-order multiplicative polynomial. In the later part of \citet{lee23}, the influence of internal dust on the FSF has been partially tested, which turned out to be adequately corrected with the former method (using the Calzetti law). However, the test has a limit, as the same attenuation curve has been used both in building the artificial spectra and in measuring them with the FSF. Thus, the dust correction test in \citet{lee23} demonstrates only that we can effectively correct dust attenuation if we already know the dust attenuation curve precisely. Unfortunately, our understanding of the real dust properties in galaxies remains incomplete, posing challenges for internal dust correction in practice.

In this paper, we investigate how the different options of internal dust correction in the FSF make a difference in the output stellar populations, using the Sydney-Australian Astronomical Observatory Multi-object Integral field spectrograph \citep[SAMI;][]{cro12} Galaxy Survey \citep{bry15} data and the Penalized PiXel-Fitting \citep[PPXF;][]{cap04,cap17,cap22} package. We test not only the total mean stellar populations (TMPs), but also the age-divided mean stellar populations (ADPs), which describe the star formation and chemical evolution history of a galaxy more extensively \citep{lee23}. Because the spectra investigated in this paper are real galaxy spectra, not artificial ones like those in \citet{lee23}, we do not know the true values in these tests, which means that we cannot measure the individual errors of the FSF results. Instead, we compare the distributions of age and metallicity as a function of galaxy stellar mass, which may be an indirect indicator of the FSF goodness: the better the FSF performs, the stronger the correlations will be.

The brief outline of this paper is as follows. Section~\ref{data} introduces the data set (SAMI) and the analysis tool (PPXF) used in this work. Section~\ref{method} describes how to compare the FSF results between different choices of dust correction. In section~\ref{result}, the mass-age relations (MTRs) and mass-metallicity relations (MZRs) are compared between the dust correction choices, for not only TMPs but also ADPs. Section~\ref{discuss} discusses the results, focusing on the advantages and limits of each dust correction choice. Finally, in Section~\ref{conclude}, the paper is concluded with some empirical guidelines for dust correction.

\section{Data and Analysis\label{data}}

We use the SAMI Galaxy Survey Data Release Three \citep{cro21}, in which the spatially-resolved spectra of 3,068 galaxies are publicly available. The target galaxies, with a redshift range of $z < 0.1$, were observed using the SAMI instrument \citep{cro12} mounted on the 3.9-m Anglo-Australian Telescope at Siding Spring Observatory in Australia. The SAMI instrument consists of 13 hexabundles, each containing 61 fibers with a $75\%$ filling factor. The angular diameter of each hexabundle's field of view is 15 arcsec, and each fiber core has a 1.6-arcsec diameter. The SAMI spectroscopy covers a wavelength range of 3700 -- 7350 {\AA}, with a blue arm (3700 -- 5700 {\AA}; R = 1730; 1.03 {\AA} per pixel) and a red arm (6250 -- 7350 {\AA}; R = 4500; 0.57 {\AA} per pixel) obtaining spectra simultaneously, resulting in a small gap between the blue and red arms.
The full width at half maximum (FWHM) for the blue and red data cubes are 2.65 {\AA} and 1.61 {\AA}, respectively \citep{van17}. During the FSF process, the blue and red spectra are merged into a single spectrum by smoothing and resampling the red spectrum, ensuring the merged spectrum maintains the same FWHM and resolution as the blue spectrum. 
The SAMI Galaxy Survey covers three regions of the Galaxy And Mass Assembly \citep[GAMA;][]{dri11} survey and eight galaxy clusters. For further details on the SAMI galaxy survey, refer to \citet{bry15} and \citet{owe17}.

For the FSF, we utilize the PPXF package with stellar population models the Medium resolution INT Library of Empirical Spectra \citep[MILES;][]{vaz10}. Adopting the Padova+00 isochrone \citep{gir00} with an initial mass function characterized by a unimodal logarithmic slope $\Gamma = 1.3$, we use the following template ages [Gyr]: [0.063, 0.079, 0.10, 0.13, 0.16, 0.20, 0.25, 0.32, 0.40, 0.50, 0.63, 0.79, 1.00, 1.26, 1.58, 2.00, 2.51, 3.16, 3.98, 5.01, 6.31, 7.94, 10.00, 12.59, 15.85], and metallicities [M/H]: [$-1.71$, $-1.31$, $-0.71$, $-0.40$, 0.00, $+0.22$].

We run the PPXF along through the following four steps:
\begin{itemize}
\item[(1)] First run with random initial guesses for velocity, velocity dispersion and noise.
\item[(2)] Second run with improved initial guesses obtained from the first run.
\item[(3)] Third run with the bad pixel masks from the second run, while setting kinematics as free parameters (as done in steps (1) and (2)), with the degree of the additive Legendre polynomial to correct the template continuum set to be 10 (\emph{degree = 10}), and without reddening correction based on a predefined attenuation curve (\emph{reddening = None}).
\item[(4)] Fourth run by fixing kinematics as obtained from the third run.
In this step, the option for dust correction is selected among three choices: [Choice-1] using the PPXF reddening correction, without any polynomial (\emph{reddening} keyword activated, \emph{degree = $-1$}, and \emph{mdegree = 0}); [Choice-2] using the multiplicative Legendre polynomial to correct the continuum shape without the PPXF reddening correction (\emph{reddening = None}, \emph{degree = $-1$}, and \emph{mdegree = 10}); and [Choice-3] without the PPXF reddening correction nor the polynomial (\emph{reddening = None}, \emph{degree = $-1$}, and \emph{mdegree = 0}). See Table~\ref{choices} for a summary.
\end{itemize}

An either additive or multiplicative Legendre polynomial can be optionally adopted in the PPXF, which aims to correct the template continuum shape during the fit, by adding or multiplying a polynomial component. Although the use of such an arbitrary polynomial may not be well justified scientifically, it is sometimes adopted for enhancing the fitting.

In this paper, we do not use the regularization (\emph{regul = 0}), but we confirmed that the overall results do not significantly change even when it is applied (\emph{regul = 5}). We use the integrated spectrum within one half-light radius in each galaxy.
The stellar masses of the sample galaxies were obtained from the SAMI input catalog, which were estimated using the $i$-band absolute magnitudes and the $g-i$ colors from photometric surveys \citep[see Sections 3.1 and 5.2 of][for details]{owe17}.

\begin{table}[t!]
\caption{Three option choices in our PPXF running\label{choices}}
\centering
\begin{tabular}{lll}
\toprule
Choice-1 & \emph{reddening} ON & polynomial OFF    \\
\midrule
Choice-2 & \emph{reddening} OFF & polynomial ON    \\
\midrule
Choice-3 & \emph{reddening} OFF & polynomial OFF    \\
\bottomrule
\end{tabular}
\tabnote{
Here, the `polynomial' indicates the multiplicative Legendre polynomial.
}
\end{table}

\section{Methods\label{method}}

Unlike \citet{lee23}, which is based on artificial spectra, we cannot determine the errors in the estimated age and metallicity compared to the true values, because we do not know them for the real galaxies. Instead, in this paper, we compare the distributions of age and metallicity as a function of stellar mass (MTR and MZR) between the three choices of the PPXF options, based on the mass-weighted mean stellar populations. In terms of statistics, it is well known that galaxies with lower stellar masses tend to be younger \citep[known as \emph{downsizing}; e.g.,][]{cow96,tre05,mag13,tom19} and more metal-poor \citep[e.g.,][]{gar87,zar94,tam01,tre04,dom23,lan23}. Thus, we expect that a better estimation of stellar populations may result in stronger correlations in MTR and MZR. In other words, the comparison of the MTRs and MZRs using the stellar populations from the three choices of the PPXF options would be an indirect method to determine which choice better reproduces the true stellar populations.

If the FSF were entirely reliable, both the total mean stellar populations across all ages (TMPs) and the mean stellar populations within a specific age interval (ADPs) would be robust. However, in reality, ADPs tend to entail greater uncertainty compared to TMPs, as they incorporate finer details of the galaxy formation history \citep{lee23}.
In other words, comparing ADPs can provide a more stringent test than comparing TMPs, aiding in determining which option choice yields the most reliable FSF results.
Therefore, our comparisons of MTRs and MZRs will involve both TMPs and ADPs. In this paper, we utilize two types of ADPs: old stellar components (stellar age $\geq 10^{9.5}$ yr $\approx 3.2$ Gyr) and young stellar components (stellar age $< 10^{9.5}$ yr). To ensure the minimum reliability of the estimated TMPs and ADPs, we restrict our analysis to datapoints with S/N $\geq 30$ for both TMPs and ADPs, as well as luminosity fraction $f_{\rm lum} \geq 30\%$ for ADPs: $f_{\rm lum}({\rm old}) \geq 30\%$ when analyzing old ADPs, while $f_{\rm lum}({\rm young}) \geq 30\%$ when analyzing young ADPs.

\section{Results\label{result}}

\begin{figure}
\centering
\includegraphics[width=85mm]{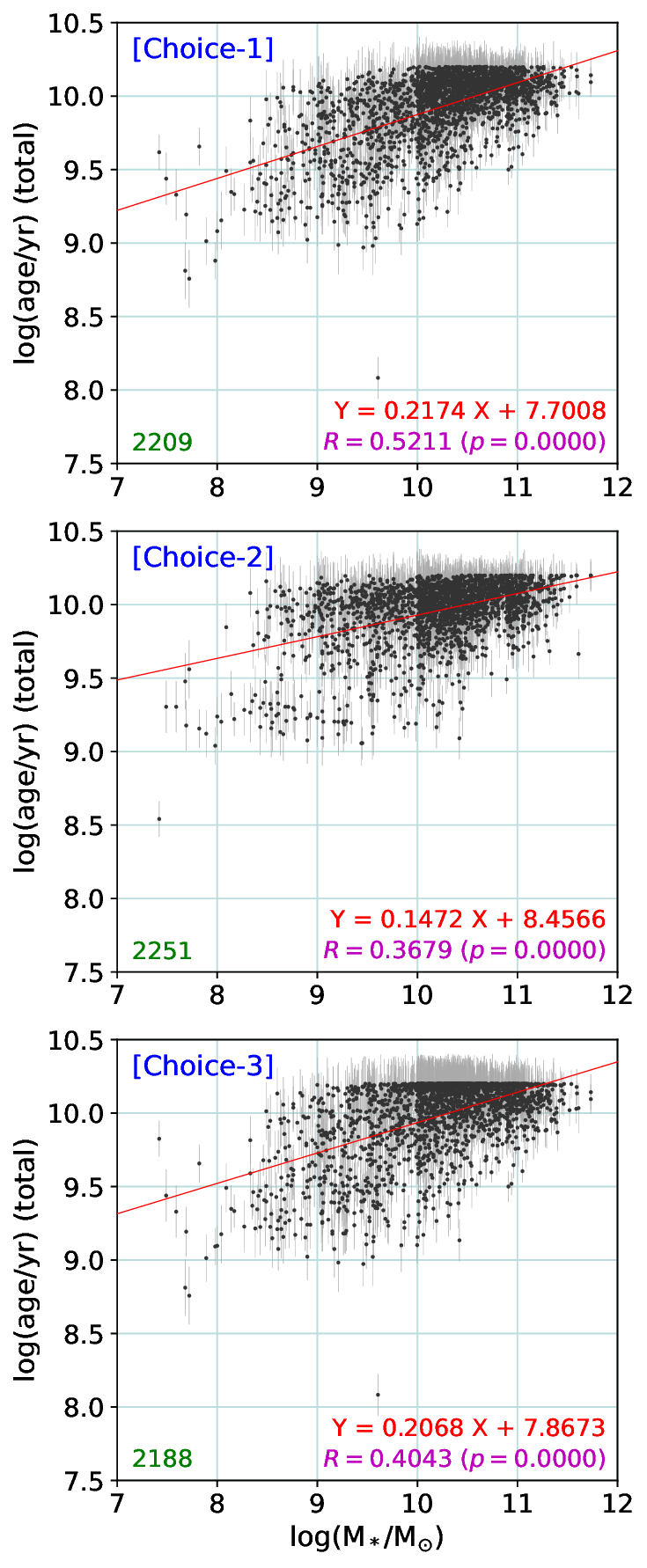}
\caption{Mass-age relations (MTRs) of the total mean stellar populations (TMPs) with respect to the three choices of PPXF dust correction options. See the text and Table~\ref{choices} for the details of the choices. The error bar of each datapoint is estimated using the uncertainty tables from \citet{lee23}. The red line shows the linear regression fit, which is also represented by a formula at the lower right corner in each panel. The Spearman's rank correlation coefficient ($R$) and p-value ($p$) are also provided. The number of datapoints that satisfy the S/N criterion is given at the left bottom of each panel.\label{mtr0}}
\end{figure}

Figure~\ref{mtr0} compares the MTRs of TMPs between the three choices of dust correction in the PPXF running. All of the three choices exhibit a moderate dependence of the TMP ages on stellar mass ($0.37\leq R \leq 0.52$, in Spearman's rank correlation coefficient), with Choice-1 showing the strongest correlation ($R=0.52$). One notable feature is the weak bimodality observed in Choice-2, reminiscent of the color bimodality seen in galaxies on the color-magnitude diagram \citep[e.g.,][]{bal04,hog04,wyd07,tay09}. Although the \emph{blue cloud} part here seems to be very poor, it may be attributed to the fact that these populations are measured within one half-light radius, primarily corresponding to the bulge region in a late-type galaxy.

\begin{figure}
\centering
\includegraphics[width=85mm]{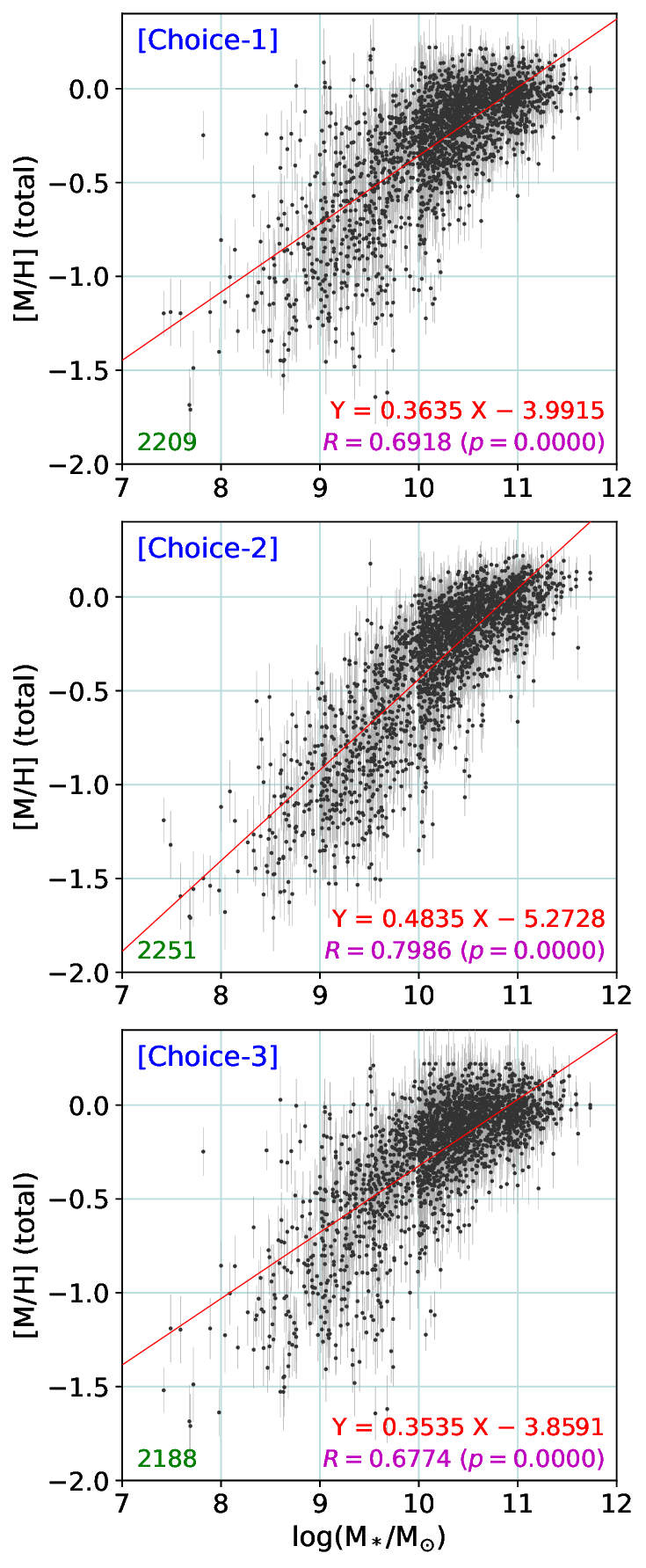}
\caption{Mass-metallicity relations (MZRs) of the TMPs with respect to the three choices of PPXF dust correction options. \label{mzr0}}
\end{figure}

In Figure~\ref{mzr0}, the MZRs of TMPs are compared. All three choices exhibit strong correlations between total metallicity and stellar mass ($R\geq0.68$), with Choice-2 showing the largest correlation coefficient ($R=0.80$) and the steepest correlation slope ($a = 0.48$). Note that the MZR may not be perfectly linear, as the slope varies depending on stellar mass. The measured slopes, derived from a simple linear regression, are predominantly influenced by densely distributed galaxies at the massive side.

\begin{figure}
\centering
\includegraphics[width=85mm]{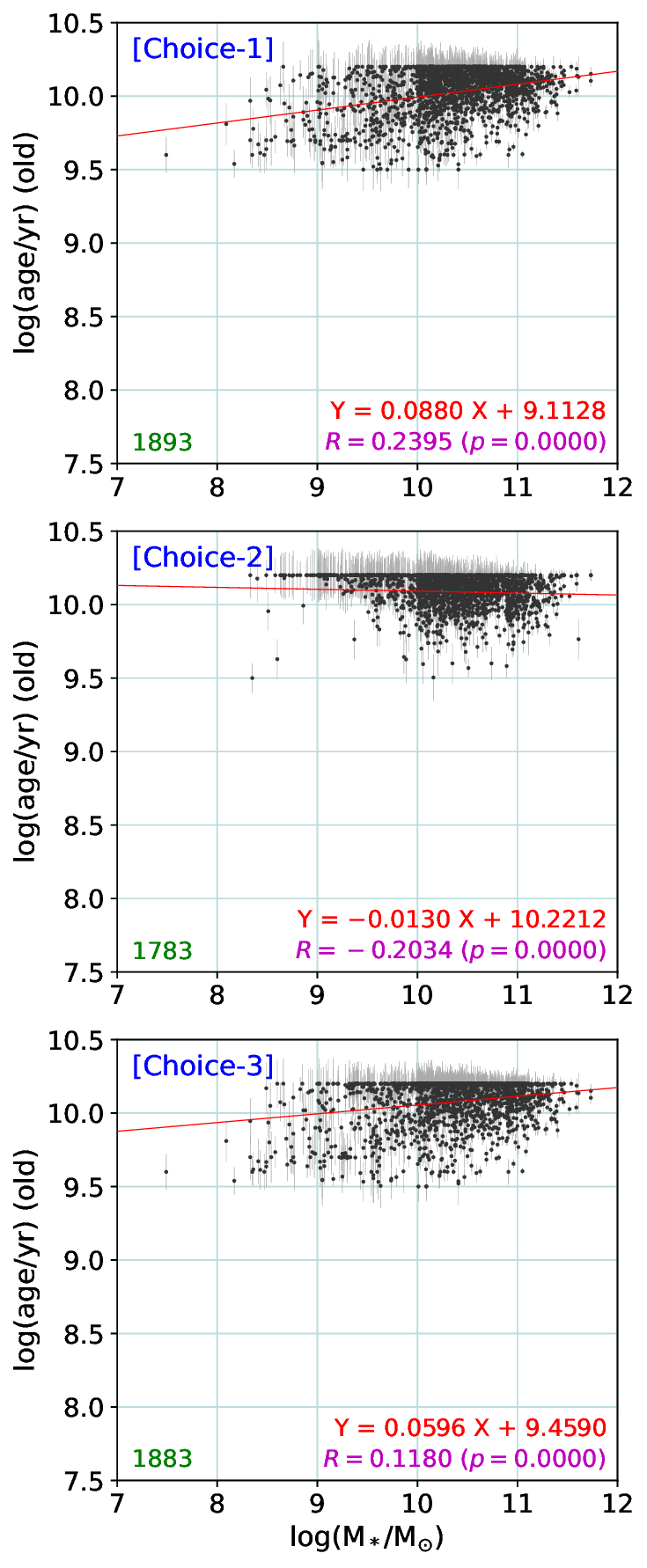}
\caption{MTRs of the old ($\geq 10^{9.5}$ yr $\approx 3.2$ Gyr) ADPs with respect to the three choices of PPXF dust correction options. The number of datapoints that satisfy the S/N and $f_{\rm lum}$ criteria is given at the left bottom of each panel.\label{mtr1}}
\end{figure}

\begin{figure}
\centering
\includegraphics[width=85mm]{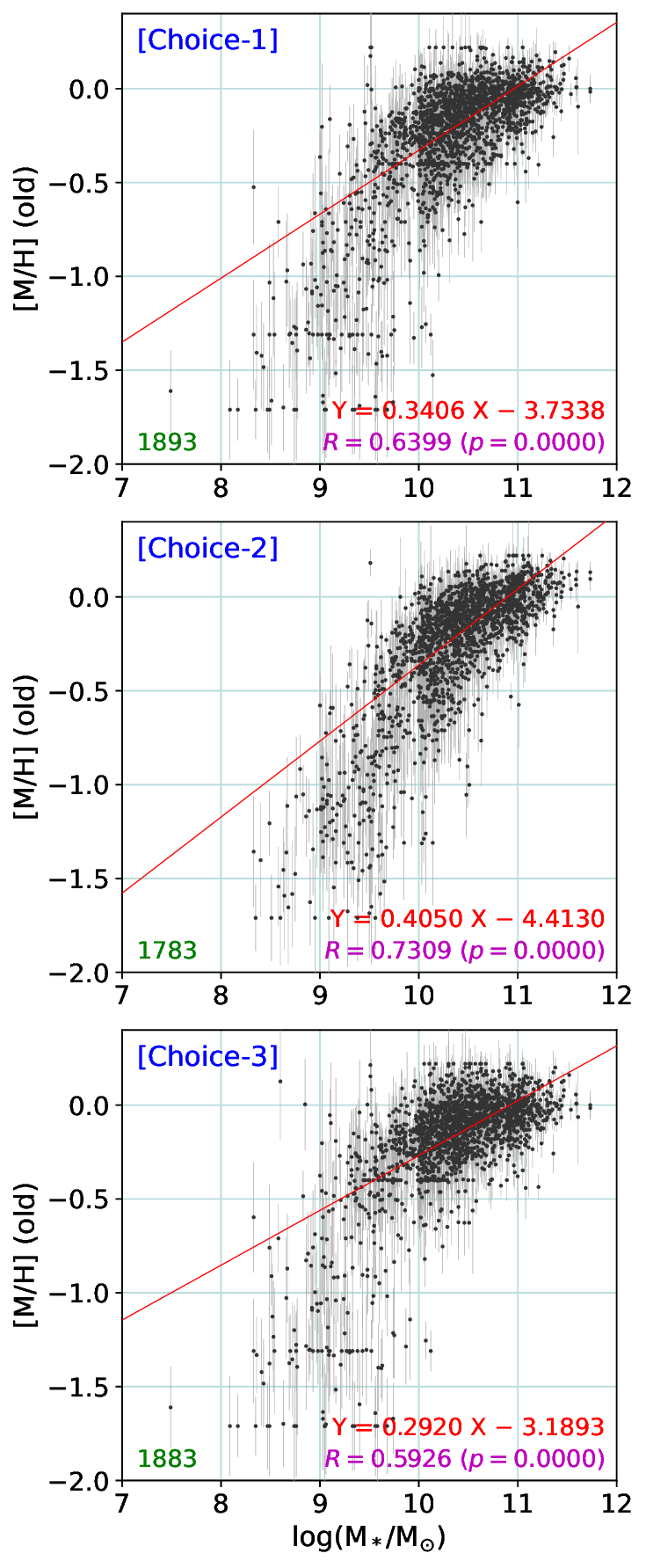}
\caption{MZRs of the old ADPs with respect to the three choices of PPXF dust correction options.\label{mzr1}}
\end{figure}

\begin{figure}
\centering
\includegraphics[width=85mm]{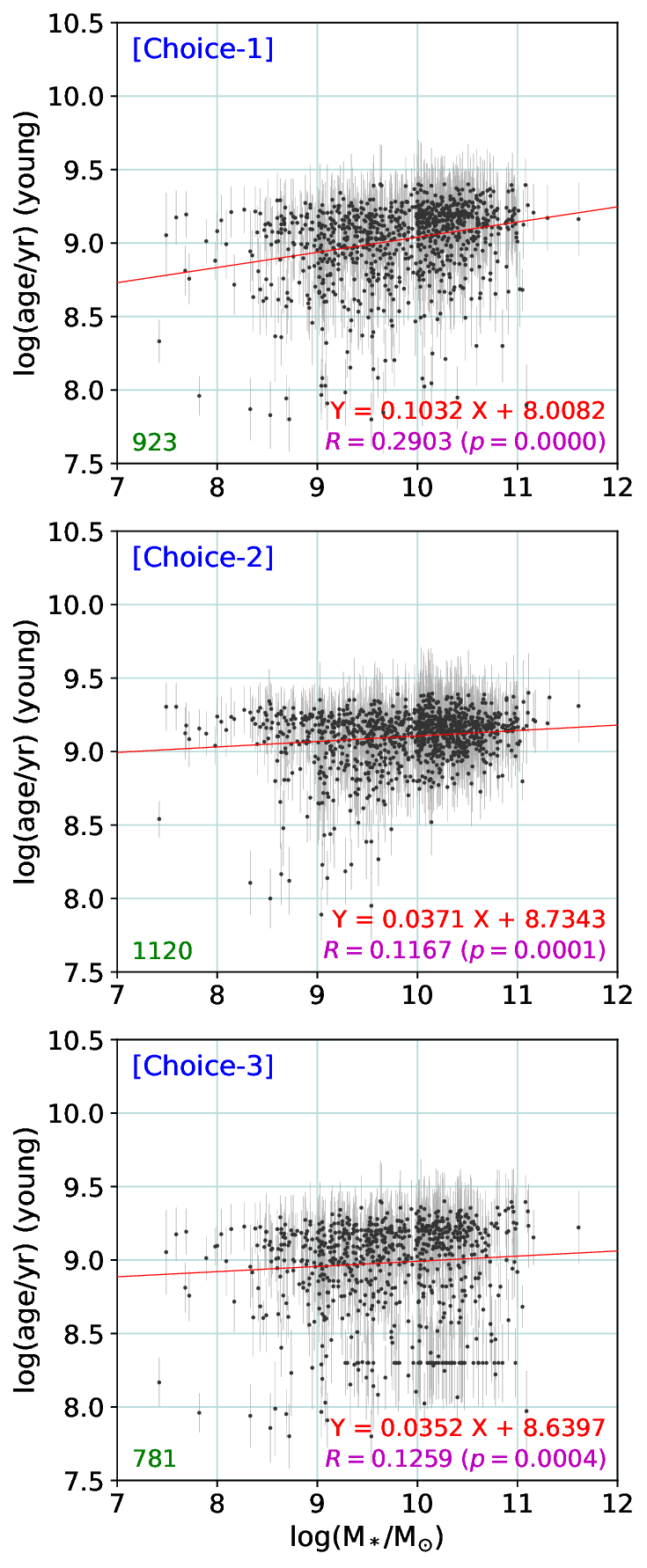}
\caption{MTRs of the young ($< 10^{9.5}$ yr) ADPs with respect to the three choices of PPXF dust correction options.\label{mtr2}}
\end{figure}

\begin{figure}
\centering
\includegraphics[width=85mm]{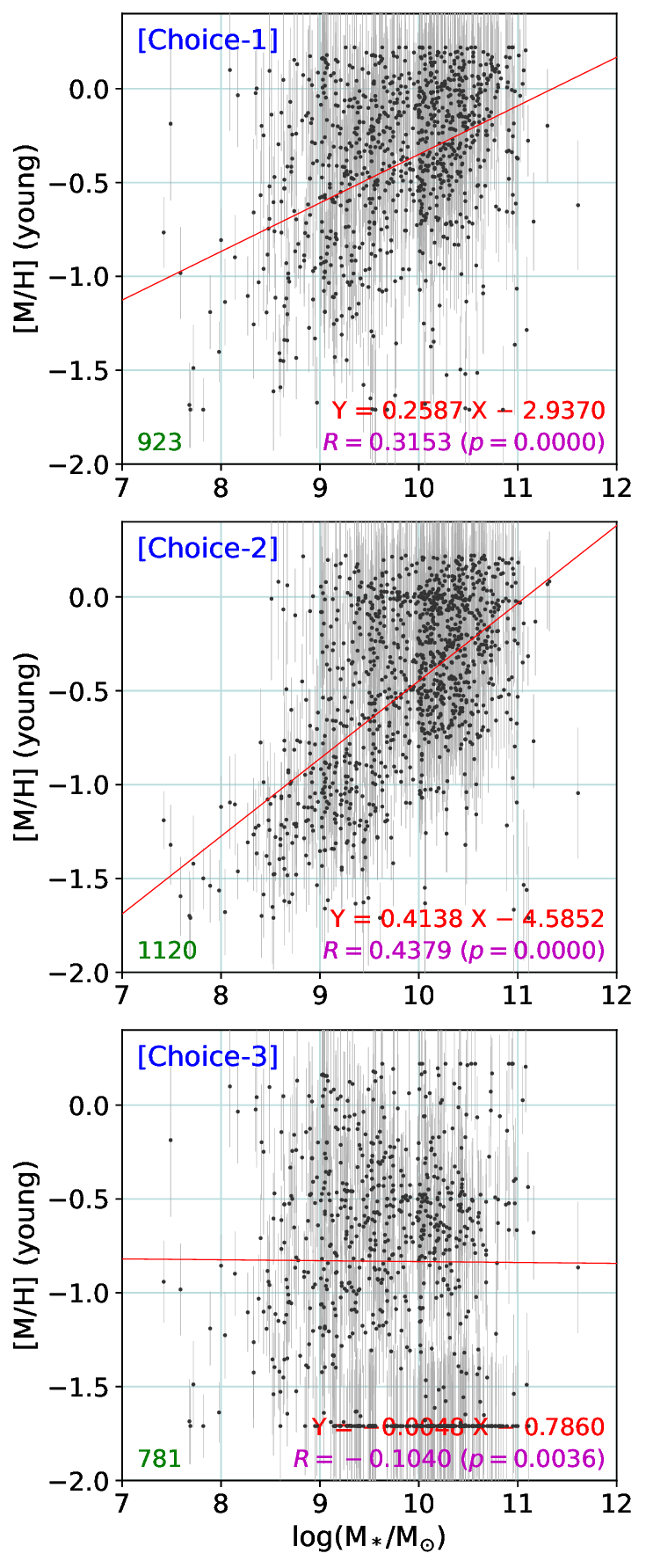}
\caption{MZRs of the young ADPs with respect to the three choices of PPXF dust correction options.\label{mzr2}}
\end{figure}

\begin{figure}
\centering
\includegraphics[width=85mm]{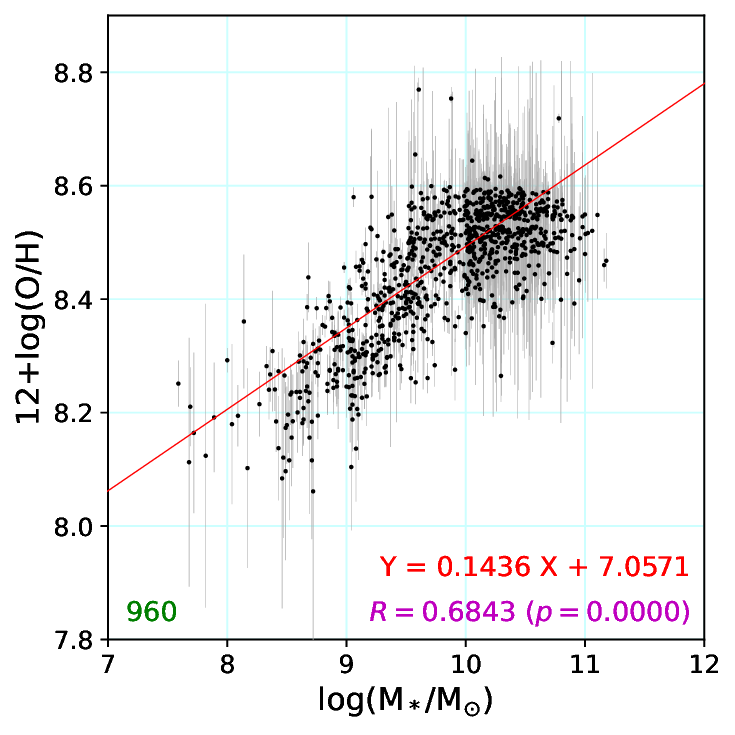}
\caption{The gas-phase MZR. The datapoints are based on those in Figure~\ref{mzr2} Choice-2, with the measurement errors of the gas-phase metallicities limited up to 0.3.\label{mzrg}}
\end{figure}

While there are variations in the tightness of MTRs and MZRs when using TMPs, they consistently demonstrate a strong dependence on stellar mass across all dust correction options. The galaxies tend to exhibit older ages and higher metallicities at higher stellar mass, aligning with previous findings. However, as mentioned in Section~\ref{method}, the FSF uncertainty tends to rise with the examination of more detailed segments in the star formation history (that is, when we use ADPs). Therefore, we further investigate MTRs and MZRs using old and young ADPs, divided by an age threshold of $10^{9.5}$ yr.

Figure~\ref{mtr1} illustrates the MTRs of old ADPs. Naturally, all datapoints are distributed above log(age/yr) = 9.5. The overall correlations are very weak irrespective of the option choices ($|R| \leq 0.24$), but a contrast emerges between the results obtained from Choice-2 versus `Choices-1 and 3'. In Choice-2, the ages tend to slightly decrease along stellar mass, with many low-mass galaxies appearing to possess extremely old ages (15.85 Gyr, the maximum age in our templates). Conversely, Choices-1 and 3 reveal MTRs with positive correlation coefficients, although they are very small, suggesting that even among the old stellar components, a tendency persists for higher mass galaxies to exhibit older ages.

In Figure~\ref{mzr1}, the MZRs of old ADPs are displayed. Similar to the MZRs of TMPs, the strongest correlation is observed in Choice-2 ($R=0.73$), which also exhibits the steepest slope ($a = 0.41$). Since the mass distribution of our sample galaxies is significantly biased to high masses ($>10^{10}$ M$_{\odot}$), the MZR slope may be primarily determined by massive galaxies rather than low-mass ones. That is, if such a mass distribution bias is appropriately corrected, the slope differences between the choices may be mitigated to some extent.

Figure~\ref{mtr2} depicts the MTRs of young ADPs. While the correlations are overall very weak, Choice-1 demonstrates a measurable correlation between age and stellar mass ($R = 0.29$). The correlations in Choices-2 and 3 are hardly discernible.

In Figure~\ref{mzr2}, the MZRs of young ADPs are depicted. Choice-2 exhibits the strongest correlation in the MZR ($R = 0.44$), while Choice-1 shows a marginal correlation ($R = 0.32$). The correlation in Choice-3 is minimal.

Finally, we demonstrate the distribution of gas-phase metallicity versus stellar mass in Figure~\ref{mzrg}, for the comparison with the MZRs of young ADPs. The gas-phase metallicity was determined from the emission line fluxes using the recipe of \citet{mar13}:
\begin{equation}
12+\log({\rm O/H}) = 8.533-0.214 \times {\rm O3N2},
\end{equation}
where O3N2 $=$ log(([O {\small III}]$_{5007}$ / H$\beta$) $\times$ (H$\alpha$ / [N {\small II}]$_{6583}$)).
The galaxies plotted in Figure~\ref{mzrg} correspond to the same sample as that in Figure~\ref{mzr2} Choice-2\footnote{Even if we adopt any other choice, the trend does not change significantly.}, which exhibits the tightest MZR of young ADPs, with limiting the measurement errors of the gas-phase metallicities up to 0.3. The gas-phase MZR demonstrates a good correlation ($R = 0.68$), but the MZR slope ($a = 0.14$) is notably shallower compared to that in Figure~\ref{mzr2} Choice-2 ($a = 0.41$). This discrepancy may arise because the young ADPs in this paper encompass ages up to approximately 3.2 Gyr, including stars that are relatively older from the perspective of recent star formation.

\section{Discussion\label{discuss}}

The key question in this paper is `which choice for dust correction is most effective for the FSF?' Unfortunately, providing a definitive answer to this question is still challenging due to our imperfect understanding of the stellar populations in real galaxies and the true nature of their dust attenuation.
In this section, we attempt to address this question by comparing the strengths and weaknesses of the three choices, based on the estimated MTRs and MZRs of TMPs and ADPs. 
The correlation coefficients for each MTR and MZR are summarized in Table~\ref{corco}.

Choice-1 (the PPXF \emph{reddening} option used) shows overall good performance, particularly in estimating MTRs. This choice presents the strongest correlations in all MTRs. It is noticeable that the MTRs of ADPs show measurable correlations ($R=0.24$ and 0.29 for old and young ADPs, respectively) only in this choice. Meanwhile, the performance of Choice-1 in estimating MZRs is also fine, but the tightness of the MZRs is always slightly worse than that in Choice-2. From the perspective of physical validity, Choice-1 may be the most reasonable choice, because it adopts a well defined (physics-based) dust attenuation curve \citep{cal00}, without use of any arbitrary polynomial. However, if the predefined dust attenuation curve cannot be fully trusted, this choice may be problematic.

\begin{table}[t!]
\caption{Correlation coefficients in all MTRs and MZRs\label{corco}}
\centering
\begin{tabular}{lrrr}
\toprule
\multicolumn{1}{c}{Choice} & \multicolumn{1}{c}{1} & \multicolumn{1}{c}{2} & \multicolumn{1}{c}{3}  \\
\midrule
MTR of TMPs & {\bf 0.521} & 0.368 & 0.404 \\
--- of old ADPs & {\bf 0.240} & $-0.203$ & 0.118 \\
--- of young ADPs & {\bf 0.290} & 0.117 & 0.126 \\
\hline
MZR of TMPs & 0.692 & {\bf 0.799} & 0.677 \\
--- of old ADPs & 0.640 & {\bf 0.731} & 0.593 \\
--- of young ADPs & 0.315 & {\bf 0.438} & $-0.104$ \\
\hline
Gas-phase MZR & \multicolumn{3}{c}{0.684} \\
\bottomrule
\end{tabular}
\tabnote{
The best correlation among the three choices is indicated in bold.
}
\end{table}

Choice-2 (the multiplicative Legendre polynomial used) demonstrates the most robust performance in estimating MZRs, as evidenced by the strongest correlations in all MZRs. On the other hand, the MTRs in this choice do not exhibit better correlations than those in the other choices. Particularly, the ages and stellar masses show no evident (or even weakly negative) correlation in the MTR of old ADPs. Regarding the MTR of TMPs, however, it is not easy to assert whether the linear correlation as shown in Choice-1 is optimal, especially considering that the TMP MTR in Choice-2 displays a weak bimodality, which may better reflect the reality \citep[e.g.,][]{bal04}. A fundamental weakness of this choice lies in the utilization of a polynomial lacking a solid physical basis. Nevertheless, given the current incomplete understanding of dust attenuation physics in galaxies, the multiplicative Legendre polynomial may offer considerable flexibility in dust correction procedures.

Choice-3 (no dust correction) seems to be the worst, because almost all MTRs and MZRs show poor correlations. This choice was tested simply on the purpose of comparison, which shows how bad the FSF results are if internal dust attenuation is not corrected at all. Nevertheless, it is noted that the MTRs of TMPs and old ADPs appear to be better than those in Choice-2.

It is still not easy to identify a single correct answer to which choice is the best for the FSF. Nevertheless, we try to draw conditional conclusions from our comparisons. First of all, we can easily reject Choice-3: for most galaxies, it is not good to ignore internal dust attenuation, especially when young stars are the primary focus.
Therefore, the real conflict lies between Choice-1 and Choice-2.
Choice-2 shows an excellent performance in estimating MZRs, whereas it is somewhat suspicious if its MTRs are realistic.
On the other hand, the MZRs in Choice-1 are overall poorer than those in Choice-2 (but still pretty good), while it presents the best correlations in the MTRs.
However, it also needs to be considered which MTR is more \emph{realistic}: a single linear correlation (Choice-1) or a bimodality (Choice-2).

\begin{figure*}
\centering
\includegraphics[width=160mm]{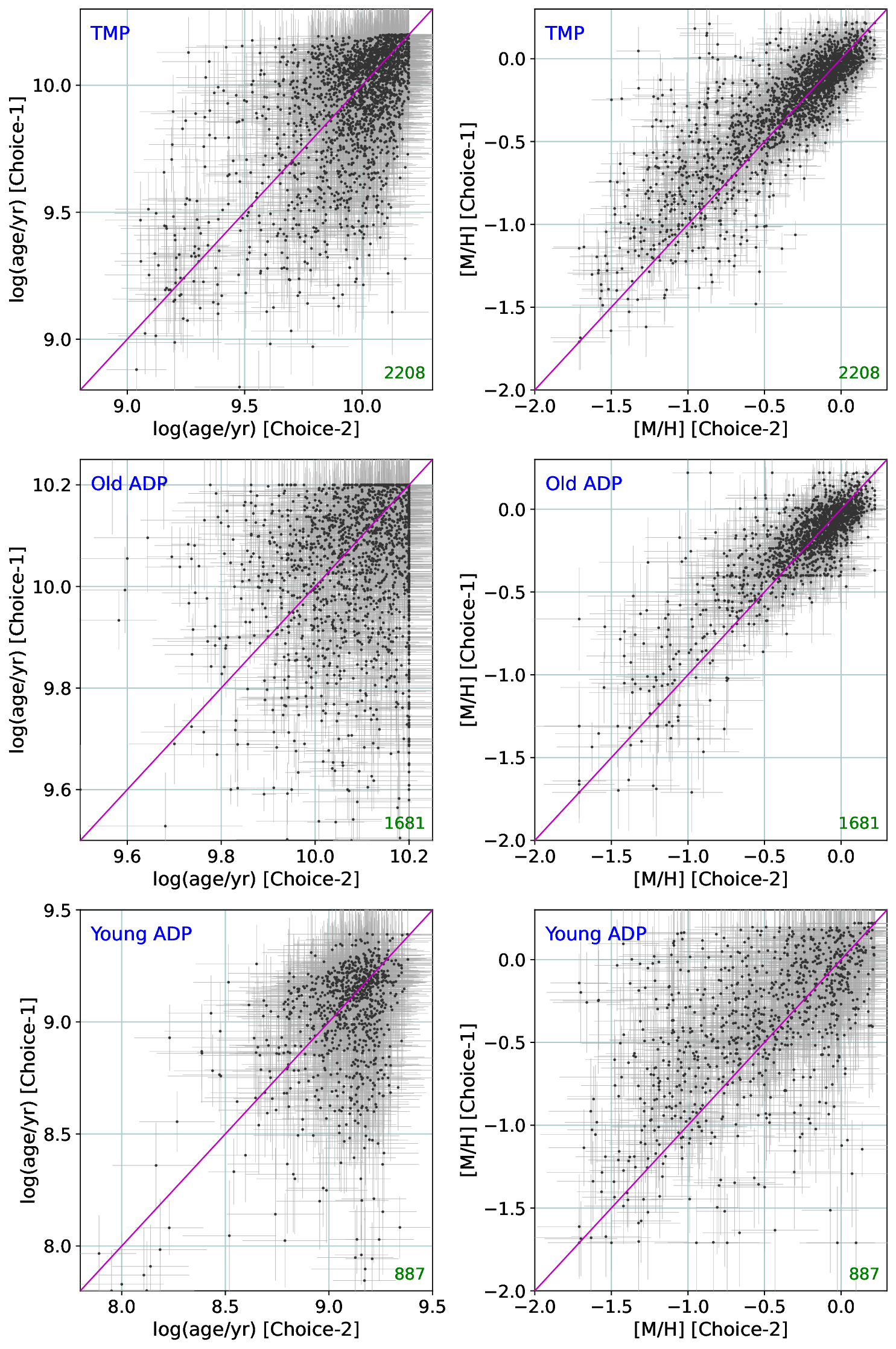}
\caption{Comparisons of age and metallicity in TMPs and ADPs between Choice-1 versus Choice-2. The magenta line indicates a one-to-one correspondence. \label{comp12}}
\end{figure*}

Figure~\ref{comp12} compares the ages and metallicities in TMPs and ADPs between Choice-1 versus Choice-2. Overall, the scatters in the age comparisons are pretty large, while the scatters in the metallicity comparisons are relatively small. It is noted that metallicity estimated in Choice-1 tends to be higher than that in Choice-2, on average, while such a bias is particularly large in young ADPs. The large scatters and notable biases, which often surpass the error bars, imply that dust correction may significantly affect the estimated stellar populations.

\section{Conclusions\label{conclude}}

We examined three choices of internal dust correction options in the FSF, using the SAMI galaxy survey data and the PPXF package, for understanding which choice returns the best results. Because we do not know the true stellar populations of the SAMI galaxies, we compared the MTRs and MZRs of their TMPs and ADPs, under the assumption that a better choice results in stronger correlations.

In this paper, we cannot pick a single best choice. Although Choice-3 may be reasonably rejected, Choices-1 and 2 have their own strengths and weaknesses at the same time. Therefore, in conclusion, we provide the following empirical guidelines for choosing between them:
\begin{enumerate}
\item[Choice-1 (activating \emph{reddening} option):]{\, \\ Stellar age is the primary concern. The MTR of galaxies is believed to show a linear correlation. Dust correction using a physics-based attenuation curve is preferred.}
\item[Choice-2 (using the multiplicative Legendre polynomial):]{\, \\ Stellar metallicity is the primary concern. The MTR of galaxies is believed to be bimodal. Flexible dust correction is preferred.}
\end{enumerate}


\acknowledgments

This work was supported by the Korea Astronomy and Space Science Institute under the R{\&}D program (Project No. 2024-1-831-01) supervised by the Ministry of Science and ICT (MSIT).
JHL and HJ acknowledge support from the National Research Foundation of Korea (NRF) grant funded by the Korea government (MSIT) (No. 2022R1A2C1004025). JC acknowledges support from the NRF grant funded by the MSIT (No. 2018R1A6A3A01013232 and 2022R1F1A1072874). SO acknowledges support from the NRF grant funded by the MSIT (No. RS-2023-00214057 and 2020R1A2C3003769).



\end{document}